**Talking to a Human as an Attitudinal Barrier: A Mixed Methods Evaluation of Stigma, Access, and the Appeal of AI Mental Health Support**

Caitlin A. Stamatis[1], Emma C. Wolfe[2], Matteo Malgaroli[3], Thomas D. Hull[1]

[1]Slingshot AI, 228 Park Ave S, PMB 679458, New York, NY 10003, US

[2]University of Virginia, Department of Psychology, 485 McCormick Rd, Charlottesville, VA 22904

[3]New York University School of Medicine, One Park Avenue, New York, NY 10016

**Abstract**

**Background:** Many individuals who could benefit from evidence-based psychotherapy do not receive it due to evaluation-sensitive concerns (e.g., shame, stigma, fear of judgment, privacy) and structural barriers (e.g., limited provider availability, long waitlists, scheduling challenges). Conversational artificial intelligence (AI) tools are increasingly used for mental health support, yet it remains unclear which therapy barriers AI helps mitigate.

**Objective:** To examine whether evaluation-sensitive (shame/stigma) and structural (access, cost/coverage) barriers to psychotherapy predict perceived helpfulness of a purpose-built AI mental health conversational tool (Ash), and whether effects differ by prior therapy experience or user engagement.

**Methods:** In a cross-sectional survey of Ash users (N=395), participants rated Ash's helpfulness (1-5) and described barriers to therapy. Open-text responses were coded for shame/stigma, access, and cost/coverage themes. Linear regressions examined associations between barriers and perceived helpfulness, adjusting for demographics and mental health covariates, with moderation by prior therapy experience. Negative binomial regressions tested whether barriers predicted platform usage.

**Results:** In the primary model (n=375), shame/stigma (B=.45, p<.001) and access barriers (B=.31, p=.020) predicted higher perceived helpfulness, whereas cost/coverage did not (B=.13, p=.262). Prior therapy experience moderated the shame effect (interaction B=.56, p=.036): shame predicted higher helpfulness among therapy-experienced users (Δ=.62, p<.001) but not therapy-naïve users (Δ=.03, p=.877). Among therapy-experienced participants comparing Ash with past therapy (n=258), shame/stigma (B=.75, p<.001) and access barriers (B=.51, p=.006) predicted rating Ash more favorably. Access barriers predicted higher engagement (IRR=1.64,

p<.001), and cost/coverage barriers predicted 70% more sessions (IRR=1.70, p<.001), whereas shame/stigma was not associated with total sessions (IRR=.80, p=.094).

**Conclusions:** AI mental health support was perceived as most helpful among users facing shame/stigma and access barriers to psychotherapy, suggesting it may function as an evaluation-safe adjunct or alternative, particularly for therapy-experienced individuals. Access and cost barriers were most predictive of usage intensity, highlighting unmet need in this population. These findings highlight the importance of aligning the design, implementation, and safety guardrails of conversational AI tools for emotional support with user-reported barriers.

**Introduction**

Mental disorders contribute substantially to global disease burden [1,2], yet many people who could benefit from psychotherapy do not receive care [3,4]. Barriers include structural constraints (e.g., provider shortages, waitlists) and evaluation-sensitive concerns such as shame, stigma, and privacy [5,6]. Conversational AI tools are increasingly used for emotional support [7,8], with nearly half of adults with a mental health condition reporting generative AI use in one recent survey [7]. Although some specialized agents show clinical promise [9–11], use of general-purpose AI for emotional support raises safety concerns including crisis detection failures, harmful advice, and privacy risks [12,13].

Despite rapid uptake, relatively little is known about why individuals turn to conversational AI for mental health support or which barriers to psychotherapy are most closely tied to perceived usefulness. Identifying these motivations has implications for harm reduction and clinical integration: if AI is attractive because it reduces interpersonal evaluation, design and messaging may need to emphasize privacy and nonjudgment; if availability is the primary driver, AI may be most useful during access gaps such as waitlists or after-hours support. These questions also matter for safeguarding users, as AI-based emotional support may be appropriate for some individuals but potentially risky in others (e.g., those experiencing or vulnerable to psychosis or mania) [14–16]. Without empirical evidence about which barriers matter and for whom, it is difficult to guide users toward safe, appropriate resources or to anticipate when AI may substitute for professional care.

In this study, we distinguish between evaluation-sensitive barriers (e.g., shame, stigma, fear of judgment) and enabling-resource barriers (e.g., access, financial and/or insurance constraints), which are among the most common obstacles to mental health care globally [17–19]. Self-presentation and stigma theories suggest that people may avoid therapy when disclosure feels identity-threatening or socially risky [20,21], particularly because psychotherapy requires sharing distress with another person in an evaluative interpersonal context [22]. Many therapeutic models argue that when processed alongside an effective psychotherapist, tolerating the distress of such vulnerability can lead to improved clinical outcomes [23,24]. However, this is only true for those who approach and/or remain in therapy despite these challenging emotions. Conversational AI may reduce these evaluation-sensitive costs by enabling more private, asynchronous disclosure without face-to-face scrutiny, leading individuals with shame-related barriers to perceive AI support as especially helpful.

Access barriers may also be associated with increased perceived helpfulness of conversational AI. Andersen's Behavioral Model of Health Services Use emphasizes the role of enabling resources—such as provider availability, time, transportation, and affordability—in shaping service use [25]. When traditional care is difficult to access because of waitlists, scheduling constraints, or financial limitations, users may turn to conversational AI for inexpensive, on-demand support. In this context, AI may be perceived as helpful not because it reduces evaluation concerns, but because it provides immediate assistance when few alternatives are available.

Together, these perspectives suggest that conversational AI may be perceived as especially helpful among individuals facing either structural access constraints or evaluation-sensitive barriers to psychotherapy. However, most widely used general-purpose large language model chatbots are not designed or evaluated as mental health interventions and may provide inconsistent or clinically inappropriate guidance. In contrast, purpose-built conversational agents developed specifically for emotional support may offer a safer, more structured alternative.

Understanding which psychotherapy barriers predict perceived helpfulness of such tools is an important step toward evaluating their role in care and informing harm-reduction strategies for users who might otherwise rely on general-purpose systems.

The need to clarify how psychotherapy barriers relate to perceived benefit from conversational AI motivated this mixed-methods study. In a sample of real-world users of an AI mental health conversational tool designed for emotional support (Ash; N=395), we examined whether qualitatively coded barriers to human psychotherapy predict perceived helpfulness of AI-based support. Based on evaluation-sensitive and enabling-resource frameworks, we hypothesized that perceived helpfulness would be higher among individuals reporting shame/stigma barriers to disclosure as well as structural barriers to accessing care (Hypothesis 1). Because perceived benefit may reflect general readiness for self-improvement rather than barrier-specific mechanisms, we further hypothesized that these associations would persist after adjusting for users' overall motivation to improve their mental health and comfort discussing emotional concerns (Hypothesis 2). Beyond perceived helpfulness, we hypothesized that users facing greater shame/stigma or structural barriers would show stronger engagement with the conversational AI platform, consistent with the possibility that AI support functions as a more accessible or lower-stakes alternative when traditional care is difficult to obtain or use (Hypothesis 3). Finally, from an exploratory perspective, we examined the relationship of shame/stigma and structural barriers to perceived helpfulness of AI mental health support relative to previous experience with a human therapist. To contextualize these quantitative patterns, we also analyzed open-ended user responses to better understand perceived advantages and limitations of AI mental health support compared with in-person therapy.

**Methods**

**Participants and Procedures**

We conducted a cross-sectional online survey of users of Ash (talktoash.com), a conversational AI tool designed specifically for emotional support, with embedded open-ended questions (mixed-methods design). The study was approved by the Institutional Review Board at NYU School of Medicine (i25-01177). The primary objective was to examine whether barriers to human psychotherapy were associated with perceived helpfulness and engagement with AI-based mental health support.

Participants were recruited from September 2024 through January 2026 from a distribution list of known Ash users and invited via email to complete a voluntary ~10-minute survey hosted on Tally. Participants were entered into a $500 lottery for survey completion. Of the 1,261 participants who completed the survey, the subsample of 395 who provided open text data were included in this study. To validate self-reported helpfulness ratings with objective behavioral data, we obtained platform usage data from Mixpanel, the analytics service tracking user engagement. We matched survey respondents to Mixpanel users via email addresses (normalized to lowercase, trimmed whitespace). All responses were de-identified prior to analysis. The final sample consisted of 395 participants who provided codable responses to at least one barrier question.

**Qualitative Coding Approach**

*Qualitative Coding and Quote Matrix:* Participants who provided open-ended responses to at least one barrier question comprised the analytic sample (n=395; 31.3% of the full sample, N=1,261). Three barrier categories were coded from the textual responses to specific questions (see "Measures" below) and used as predictors in quantitative analyses: (1) Shame/stigma barriers (e.g., embarrassment, fear of judgment, stigma, difficulty admitting need for help); (2)

Access barriers (e.g., provider availability, waitlists, scheduling constraints, geographic limitations, difficulty finding appropriate fit); (3) Cost/coverage barriers (e.g., affordability, insurance coverage). Each category was coded as present (1) or absent (0) at the participant level; a barrier was considered present if it appeared in either open-ended response. To support transparency and contextual interpretation, we compiled a quote matrix linking themes and subthemes to exemplar quotations (Supplementary Tables 5 & 6) Two independent coders (one a licensed clinical psychologist, and one an advanced clinical psychology doctoral student) iteratively coded all responses, and discrepancies were resolved through discussion to consensus [26].

*Exploratory Coding of Perceived Limitations of Ash:* We conducted an exploratory thematic review of open-ended responses describing perceived limitations of Ash or situations in which participants preferred a human mental health professional (n=42). Responses were inductively coded to identify recurring themes (e.g., missing human connection, limited contextual understanding, insufficient challenge, overly affirming tone, lack of focus, or memory errors). In line with the mixed methodology of concurrent triangulation [27], analysis was intended to contextualize quantitative findings rather than test hypotheses.

**Measures**

*Barriers to Traditional Therapy (Primary Predictors):* Barriers to psychotherapy were elicited using two open-ended questions. Participants who had never spoken with a mental health professional were asked: "If you've never seen a mental health professional, why not?" Participants with prior therapy experience were asked: "Please use the space below to share any other thoughts about your experience with mental health professionals."

*Perceived Helpfulness of Ash (Primary Outcome):* Participants rated the extent to which they believed Ash could help with their mental health on a 5-point Likert scale ranging from 1 ("Not at all") to 5 ("A great deal"). This served as the primary outcome for Hypotheses 1 and 2.

*Control Variables:* Comfort discussing mental health was assessed using a 5-point item ("Very uncomfortable" to "Very comfortable"). Motivation to improve mental health was assessed with a 5-point item ("Not important" to "Top priority"). Demographic covariates included age ($<35$, 35-54, $\geq55$), gender (man, woman, nonbinary/other), and prior therapy experience (yes/no).

*Comparative Helpfulness Relative to Past Therapy (Exploratory Outcome):* Participants with prior therapy experience rated: "How would you compare your experience with Ash to your previous experience with a mental health professional?" on a 5-point scale (1=Much less helpful to 5=Much more helpful). For exploratory binary analyses, responses were dichotomized as Ash rated more helpful than past therapy (Somewhat more helpful or much more helpful; 4-5) versus Ash rated the same or less helpful (1-3). "Not sure" responses were excluded from these models.

*Engagement Outcomes (Behavioral Validation):* The primary engagement metric was total sessions, defined as discrete instances of platform use recorded by Mixpanel from September 2024 through January 2026.

**Analytic Strategy**

For all models, statistical significance was evaluated at $\alpha=.05$ (two-tailed). We report unstandardized regression coefficients (B) for linear models, odds ratios (ORs) for ordinal or binary models, and partial $R^2$ as an effect-size index for continuous outcomes.

*Hypotheses 1 and 2: Barriers and Perceived Helpfulness*: We used linear regression to test whether shame/stigma, access, and cost/coverage barriers predicted perceived helpfulness of Ash (Hypothesis 1), controlling for age, gender, and comfort discussing mental health. To test whether associations persisted beyond general motivation to improve one's mental health, we added importance of improving mental health as a covariate (Hypothesis 2). To test whether prior therapy experience moderated evaluation-sensitive barrier effects, we added prior therapy experience and a Shame/Stigma × Prior Therapy Experience interaction term into the model. Significant interactions were probed using simple slopes.

*Hypothesis 3: Barriers and Engagement:* We used negative binomial regression to model total sessions as a function of the three barrier types, controlling for demographics and mental health importance. This approach was selected due to overdispersion in session counts.

*Exploratory Analyses:* Among participants with prior therapy experience, we examined whether barriers predicted ratings of Ash relative to past using linear regression, controlling for age, gender, comfort discussing mental health, and importance of improving mental health.

*Sensitivity and Robustness Analyses*: To assess the robustness of primary findings, we conducted several sensitivity analyses. First, because perceived helpfulness was measured on an ordinal Likert scale, we re-estimated primary models using ordinal logistic regression, treating helpfulness as an ordered categorical outcome. Second, to examine convergent validity of the qualitative shame/stigma coding, we tested whether the shame/stigma barrier indicator predicted lower self-reported comfort discussing mental health with others. For exploratory analyses comparing Ash with prior therapy among participants with therapy experience, we additionally re-estimated models using a binary outcome indicating whether Ash was rated as more helpful than past therapy (ratings of 4-5) versus the same or less helpful (ratings of 1-3), excluding "Not sure" responses. All sensitivity analyses used the same covariates as primary models. Rates of missing data were low (0-2% per variable), and missing data were handled using listwise deletion.

<div align="center">

**Results**

</div>

**Sample Characteristics**

Of the full sample (N=1,261), 395 participants (31.3%) provided codable open-text responses and comprised the analytic sample. Analytic sample participants were predominantly female (62.9%), with nearly half aged 35-54. Most had prior therapy experience (73.9%). Mean perceived helpfulness of Ash was 3.95 (SD=0.98), with 70.8% rating Ash as quite a bit or a great deal helpful. See Supplementary Materials for details on comparisons of the analytic sample to excluded participants (Supplementary Table S1) and details on engagement data descriptives.

**Table 1.** Characteristics of Analytic Sample (n=395)[a]

| Characteristic | n (%) or M (SD) |
|---|---|
| **Age** | |
| 18-24 | 23 (5.8%) |
| 25-34 | 82 (20.8%) |
| 35-44 | 117 (29.6%) |
| 45-54 | 80 (20.3%) |
| 55-64 | 64 (16.2%) |
| 65-74 | 22 (5.6%) |
| 75+ | 3 (0.8%) |
| **Gender** | |
| Man | 141 (35.8%) |
| Woman | 248 (62.9%) |
| Nonbinary/Other | 5 (1.3%) |
| Missing | 1 (0.3%) |

| Race/ethnicity | |
|---|---|
| White | 321 (81.3%) |
| Asian | 32 (8.1%) |
| Hispanic/Latino | 29 (7.3%) |
| Black/African-American | 18 (4.6%) |
| Native American/Alaska Native | 6 (1.5%) |
| Middle Eastern/North African | 6 (1.5%) |
| Pacific Islander | 5 (1.3%) |
| Missing | 2 (0.5%) |
| **Ever spoken to mental health professional** | |
| Yes | 291 (73.7%) |
| No | 102 (25.8%) |
| Missing | 2 (0.5%) |
| **Ash helpfulness (1-5 scale)** | |
| Mean (SD) | 3.95 (0.98) |
| Positive rating (4-5) | 274 (70.8%) |
| **Comfort discussing mental health (1-5 scale)** | |
| Mean (SD) | 3.25 (1.34) |
| **Importance of improving mental health (1-5 scale)** | |
| Mean (SD) | 4.20 (0.87) |
| **Barriers identified (qualitatively coded)** | |
| Shame/stigma | 74 (18.8%) |
| Access | 69 (17.5%) |
| Cost/coverage | 96 (24.3%) |

[a]This table describes the 395 participants (31.3% of total sample, N=1,261) who provided written responses to open-ended barrier questions that could be qualitatively coded.

## Hypothesis 1: Psychotherapy Barriers and Perceived Helpfulness of Mental Health AI

Shame/stigma and access barriers were associated with higher perceived helpfulness of Ash (Table 2, Unadjusted Values; Figure 1, Figure 2). Reporting shame/stigma barriers was associated with a 0.45-point increase in helpfulness on the 1-5 scale (p<.001), accounting for 3.3% of unique variance (partial R²=.033; Cohen's d=0.42). Access barriers were associated with a 0.31-point increase (p=.020; partial R²=.015; d=0.29). Cost/coverage barriers were not significantly associated with helpfulness (*p*=.262; partial R²=.003). Results replicated when treating helpfulness as an ordered categorical variable (ordinal logistic regression): shame/stigma (OR=2.26, *p*<.001), access (OR=1.66, *p*=.032), cost (OR=1.26, *p*=.323). (2) The coded shame/stigma barrier variable significantly predicted lower comfort discussing mental health with others (B=-0.14, *p*<.01), suggesting convergent validity.

**Table 2.** Barriers Predicting Perceived Helpfulness of Mental Health AI: Unadjusted And Adjusted For Mental Health Importance

| Predictor[a] | **Unadjusted Values** | | | | | | **Adjusted Values** | | | | | |
|---|---|---|---|---|---|---|---|---|---|---|---|---|
| | **B** | **SE** | **95% CI** | **p** | **Partial R²** | **Sig** | **B** | **SE** | **95% CI** | **p** | **Partial R²** | **Sig** |
| **Shame/Stigma barriers** | **0.45** | **0.13** | **[0.20, 0.71]** | **<.001** | **.033** | **\*\*\*** | **0.42** | **0.12** | **[0.18, 0.67]** | **<.001** | **.031** | **\*\*\*** |
| **Access barriers** | **0.31** | **0.13** | **[0.05, 0.57]** | **.020** | **.015** | **\*** | **0.25** | **0.13** | **[0.01, 0.50]** | **.045** | **.011** | **\*** |
| Cost barriers | 0.13 | 0.12 | [-0.10, 0.36] | .262 | .003 | ns | 0.12 | 0.11 | [-0.10, 0.35] | .273 | .003 | ns |
| **MH importance** | — | — | — | — | — | — | **0.34** | **0.06** | **[0.23, 0.45]** | **<.001** | **.094** | **\*\*\*** |
| Comfort discussing MH | 0.03 | 0.04 | [-0.04, 0.10] | .440 | .002 | ns | -0.00 | 0.04 | [-0.07, 0.07] | .947 | .000 | ns |
| Age 35-54 (vs <35) | -0.11 | 0.12 | [-0.35, 0.12] | .354 | .002 | ns | -0.13 | 0.12 | [-0.36, 0.09] | .249 | .004 | ns |
| Age 55+ (vs <35) | 0.08 | 0.14 | [-0.20, 0.36] | .584 | .001 | ns | 0.11 | 0.14 | [-0.16, 0.38] | .417 | .002 | ns |
| Woman (vs Man) | 0.07 | 0.11 | [-0.14, 0.27] | .538 | .001 | ns | 0.01 | 0.10 | [-0.18, 0.21] | .894 | .000 | ns |
| Nonbinary (vs Man) | -0.31 | 0.45 | [-1.18, 0.57] | .491 | .001 | ns | -0.36 | 0.42 | [-1.19, 0.48] | .403 | .002 | ns |

[a]Model: n=375, R²=.05; Adjusted R²=.036, F(8,366)=2.76, p<.001

**Figure 1.** Regression coefficients for each of the 3 barrier types in a model predicting perceived helpfulness of AI therapy.

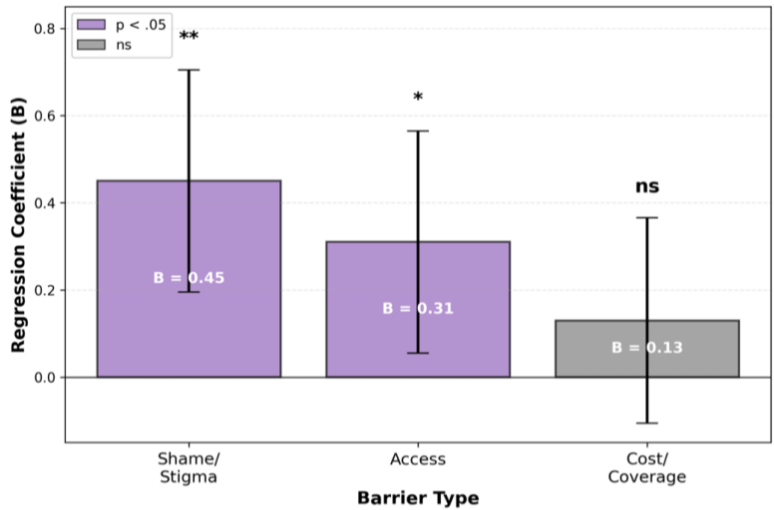

Error bars represent 95% confidence intervals. *p<.05, **p<.001, ns=not significant (p>.05).
Model 1: n=375, R²=.057, Adjusted R²=.036, p<.001. Controls: comfort, age, gender.

**Figure 2.** Helpfulness ratings for those who endorsed ("Yes") vs. did not endorse ("No") each of the 3 coded barriers.

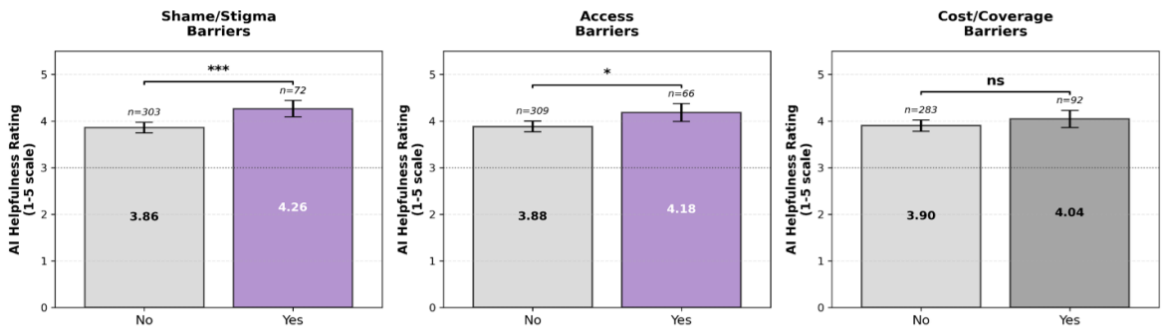

Error bars represent 95% CIs. ***p<.001, *p<.05, ns=not significant. Dotted line = Neutral (3).
Higher values = more helpful. n=375. Controls: comfort, age, gender.

## Hypothesis 2: Robustness to Motivation

After adjusting for the importance of improving mental health, shame/stigma (B=0.42, p<.001; partial R²=.031) and access barriers (B=0.25, p=.045; partial R²=.011) remained significant predictors of perceived helpfulness (Table 2; Adjusted Values), indicating that these associations were not attributable solely to general motivation or perceived need.

*Moderation by Prior Therapy Experience*

Prior therapy experience moderated the association between shame/stigma barriers and perceived helpfulness (interaction B=0.56, p=.036; partial R²=.012; Table 3). Simple slopes indicated that shame/stigma barriers were associated with higher perceived helpfulness among therapy-experienced participants (Δ=0.62, p<.001; Cohen's d=0.64; Supplementary Table S2; Supplementary Table S3; Figure 3), but not among therapy-naïve participants (Δ=0.03, p=.877).

**Table 3.** Regression Model Testing Shame/Stigma × Prior Therapy Experience Moderation

| Predictor[a] | B | SE | 95% CI | p | Partial R² |
|---|---|---|---|---|---|
| Stigma/Shame Barriers | 0.05 | 0.22 | [-0.37, 0.47] | .821 | .000 |
| Access Barriers | 0.28 | 0.13 | [0.03, 0.53] | .028 | .013 |
| Cost Barriers | 0.07 | 0.12 | [-0.17, 0.32] | .547 | .001 |
| Comfort discussing MH | 0.01 | 0.04 | [-0.07, 0.08] | .892 | .000 |
| MH Importance | 0.33 | 0.06 | [0.22, 0.44] | <.001 | .089 |
| Therapy Experience | -0.21 | 0.14 | [-0.48, 0.07] | .144 | .006 |
| **Shame × Therapy** | **0.56** | **0.26** | **[0.04, 1.07]** | **.036** | **.012** |
| Age 35-54 (vs <35) | -0.14 | 0.12 | [-0.37, 0.09] | .221 | .004 |
| Age 55+ (vs <35) | 0.12 | 0.14 | [-0.15, 0.39] | .394 | .002 |
| Woman (vs Man) | 0.00 | 0.10 | [-0.20, 0.20] | .974 | .000 |
| Nonbinary (vs Man) | -0.27 | 0.43 | [-1.11, 0.56] | .520 | .001 |

[a]R²=.157, Adjusted R²=.131, F(11,360)=6.09, p<.001

**Figure 3.** Perceived Helpfulness of Ash by Shame/Stigma Barrier and Prior Therapy Experience

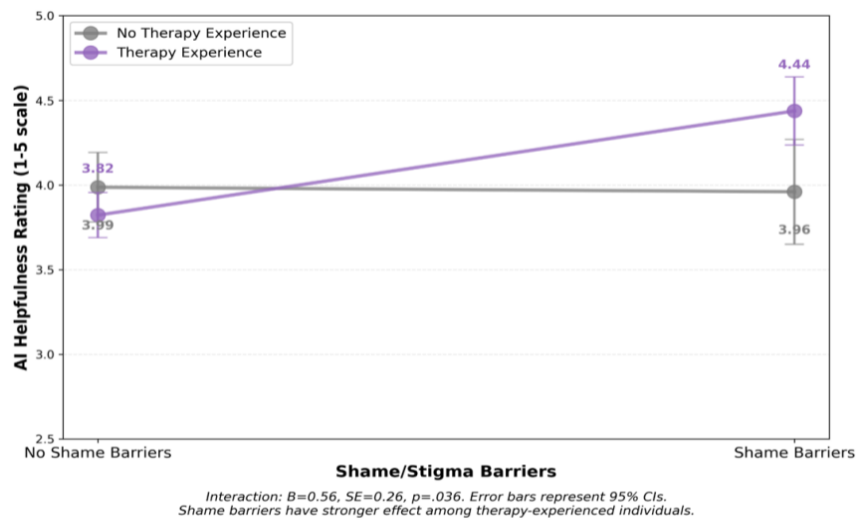

Interaction: B=0.56, SE=0.26, p=.036. Error bars represent 95% CIs.
Shame barriers have stronger effect among therapy-experienced individuals.

**Hypothesis 3: Barriers and Engagement (Behavioral Validation)**

Access barriers (IRR=1.64, p<.001) and cost/coverage barriers (IRR=1.70, p<.001) predicted greater platform engagement, corresponding to 64% and 70% more sessions, respectively (Table 4). Shame/stigma barriers were not associated with total sessions (IRR=0.80, p=.094), despite strongly predicting perceived helpfulness. Figure 4 displays mean total sessions by barrier type. Notably, 77.7% of the analytic sample engaged with the platform at least once, and 31.9% became Power Users with 31 or more sessions, indicating substantial sustained engagement beyond initial trial.

**Table 4.** Barriers Predicting Behavioral Engagement with Mental Health AI

| Predictor[a] | B | SE | IRR[b] | 95% CI | p |
|---|---|---|---|---|---|
| Shame/stigma barriers | -0.229 | 0.137 | 0.80 | [0.61, 1.04] | 0.094 |
| Access barriers | 0.494 | 0.139 | 1.64 | [1.25, 2.15] | <.001 |
| Cost/coverage barriers | 0.529 | 0.123 | 1.70 | [1.33, 2.16] | <.001 |
| Comfort discussing MH[c] | 0.023 | 0.040 | 1.02 | [0.95, 1.11] | 0.571 |
| Age <35 (ref: 35-54) | -0.240 | 0.127 | 0.79 | [0.61, 1.01] | 0.058 |

| | 0.215 | 0.133 | 1.24 | [0.95, 1.61] | 0.107 |
| Age 55+ (ref: 35-54) | | | | | |
| | -0.026 | 0.110 | 0.97 | [0.78, 1.21] | 0.815 |
| Woman (ref: man) | | | | | |
| | 0.223 | 0.467 | 1.25 | [0.50, 3.12] | 0.633 |
| Nonbinary (ref: man) | | | | | |
| MH importance | 0.084 | 0.061 | 1.09 | [0.97, 1.23] | 0.168 |

[a]Model used negative binomial distribution with log link function to account for overdispersion, Pseudo R² (Cox-Snell): 0.127
[b]IRR = Incidence Rate Ratio (exponentiated coefficient).
[c]MH = mental health.

**Figure 4.** Mean total sessions for those reporting each barrier to traditional therapy ("Present") vs. those not reporting that barrier ("Absent").

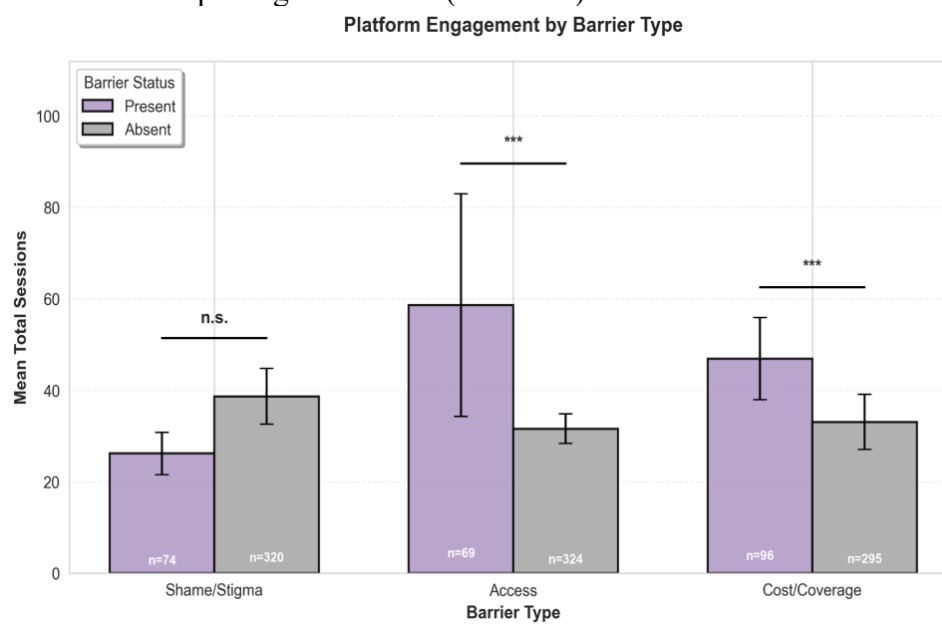

## Exploratory: Comparison With Past Therapy

Among participants with prior therapy experience who provided comparison ratings (n=258), shame/stigma (B=0.75, p<.001) and access barriers (B=0.51, p=.006) predicted rating Ash as more helpful than prior therapy (Table 5; Figure 5). Cost/coverage barriers showed no association. Binary sensitivity analyses yielded consistent results (Supplementary Table S4). In this subsample, mean rating on this item was 3.88 (SD=1.20), with 67.1% rating AI as more helpful than past therapy (4 or 5 on scale).

**Table 5.** Linear Regression: Predicting Ratings of Ash Relative to Past Human Therapy (Continuous)

| Predictor[a] | B | SE | 95% CI | p | Partial R² |
|---|---|---|---|---|---|
| Stigma/Shame Barriers | 0.75 | 0.19 | [0.38, 1.12] | <.001 | .061 |
| Access Barriers | 0.51 | 0.18 | [0.15, 0.87] | .006 | .030 |
| Cost Barriers | 0.15 | 0.21 | [-0.27, 0.56] | .488 | .002 |
| MH Importance | 0.33 | 0.08 | [0.17, 0.50] | <.001 | .059 |
| Comfort discussing MH | -0.08 | 0.06 | [-0.19, 0.03] | .140 | .009 |
| Age 35-54 (vs <35) | -0.11 | 0.17 | [-0.45, 0.23] | .536 | .002 |
| Age 55+ (vs <35) | 0.12 | 0.20 | [-0.27, 0.50] | .561 | .001 |
| Woman (vs Man) | 0.17 | 0.15 | [-0.12, 0.46] | .257 | .005 |
| Nonbinary (vs Man) | -0.09 | 0.66 | [-1.39, 1.20] | .888 | .000 |

[a]Outcome: Rating AI relative to past therapy on continuous 1-5 scale (1=Much less helpful to 5=Much more helpful. Model fit: R²=.177, Adjusted R²=.147, F(9,248)=5.93, p<.001, Reference categories: Age <35, Gender = Man.

**Figure 5.** Mean rating of AI relative to past human therapy for those reporting each barrier to traditional therapy ("Present") vs. those not reporting that barrier ("Absent").

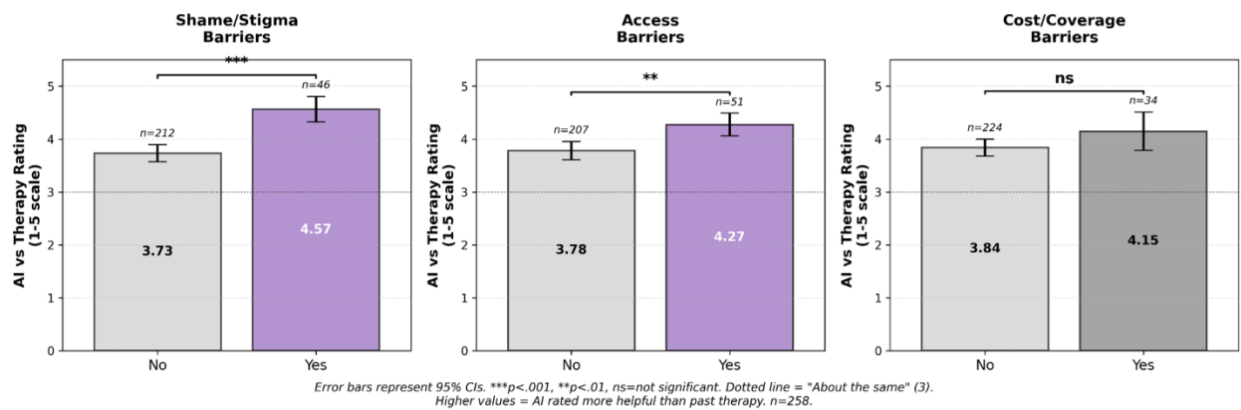

## Qualitative Findings

*Barriers and Reasons for Preferring Ash*

Across open-ended responses, participants described turning to Ash for (1) evaluation-safe disclosure (e.g., reduced shame, stigma, and fear of judgment), (2) immediate or as-needed access when traditional therapy felt logistically unavailable, and (3) as an alternative after negative or mistrust-inducing experiences with therapists. Although cost/coverage concerns were common, they did not predict perceived helpfulness in the quantitative models, suggesting that affordability may be more closely tied to reach and adoption than to perceived fit among current users.

In addition to the three barrier categories used in regression models, several recurring subthemes emerged in the quote matrix, including concerns about privacy and confidentiality, difficulty finding a therapist who felt like a good relational fit, skepticism about the authenticity or incentives of the therapeutic relationship, and perceptions that prior therapy experiences were dismissive or harmful. Representative quotations are provided in Table 6, with additional examples in Supplementary Table 5

**Table 6.** Exemplar quotations illustrating barriers to traditional psychotherapy and perceived advantages of AI mental health support

| Barrier | Quote |
|---|---|
| Prior negative experiences with therapist | "The vast majority [of therapists] I've seen haven't been helpful & some even harmful…because of these situations, I lack trust…which is why I turned to AI." |
| Trust/comfort | "Firstly, I'm terrified of the idea of opening up to someone real who talks to me for profit…" |
| Shame/stigma | "I'm embarrassed. Guys should have it together. We aren't supposed to talk about how we feel and what is killing us." |
| Fear of judgment/ misunderstanding | With Ash, I feel like I can unpack difficult feelings or conflicting thoughts honestly - including sensitive topics about identity, belonging, or self-perception - without worrying about being misunderstood or seen in a particular way." |

| Privacy/anonymity | "I simple do not feel comfortable discussing my personal life with someone who I would have to meet regularly. I prefer keeping it anonymous, which is why I choose an AI chat bot." |
| Limited availability and waitlists | "It's hard to get appointments – [the therapists'] calendars are most commonly full and they do not accept new patients." |
| Need for on-demand support | "The specialists in our country are booked months ahead. Also, it does not seem logical that I cannot discuss the burning questions I have as soon as possible. They lose relevance when I have to wait several days or weeks." |
| Skeptical of benefit from therapy | "It felt like [therapists] were all just telling me everything that I already know - like breath work and meditation and journaling - which I do all of those things. I don't need another 10-step program that doesn't really fit everything I'm dealing with.." |
| Cost/Coverage | "I am on a very fixed in income and I am extremely grateful that I can use Ash for free, if I was required to pay for the services I don't think I would be able to continue using it and it really does help me with my mental health issues." |
| | "I'm on Medicare and MH practitioners are almost impossible to locate on your advantage program. I've been on waiting list a couple times and never heard back." |

## Qualitative Findings: Perceived Limitations of Ash Relative to Human Therapy

Among participants who provided critical feedback about Ash (n=42), the most common limitations related to the absence of a human relationship, limited ability to read contextual or nonverbal cues, and concerns that the AI was insufficiently challenging, overly affirming, or inconsistently focused. These qualitative critiques help contextualize when AI mental health support may be perceived as insufficient and underscore the importance of clear use-case boundaries and expectations. Representative quotations and percentages are provided in Table 7; additional examples are available in Supplementary Table 6.

**Table 7.** Exemplar quotations illustrating perceived limitations of Ash and reasons to prefer a human professional

| Issue | Quote | Percentage (n=42) |
|-------|-------|-------------------|
| Missing human element | "The thing about a mental health professional is…them being a real human being…they are more present, and more empathetic, able to read your expressions and emotions, and give you their full attention." | 40.5% (17/42) |
| Cannot read contextual cues | "People can read micro expressions, changes in voice, delays in appointments.. etc. that the AI cannot." | 28.6% (12/42) |
| Does not appropriately provide and/or apply insight | "Ash is less helpful in providing insight but more helpful in providing thought provoking and challenging questions about the topics I specifically want to talk about." | 26.2% (11/42) |
| Does not challenge beliefs or provide new perspectives | "What I've realized about Ash (my husband and I used it for the same issue) is that AI always sides with the person using the app. There's no "challenge". An actual therapist, in the past, has challenged my viewpoint." | 19.0% (8/42) |
| Lacks focus or direction | "I think Ash has less of an agenda, which sometimes means that it's less focused and has less of a direction. I thought talking to Ash was insightful in a single conversation, but I don't know that it would grow the way therapy does across multiple sessions." | 19.0% (8/42) |
| Mechanical/ repetitive | "I find AI to be creepy. Asking basic questions, gathering info that's a good thing. Having a machine interact with me on a personal level, again, creepy." | 14.3% (6/42) |
| Overly positive/affirming | "Less sycophantic responses, I don't want a yes man…" | 11.9% (5/42) |
| Errors/ misremembering | "[Ash] didn't remember who I was or some details at times." | 9.5% (4/42) |

## Discussion

### Principal Findings

In this survey of users of a purpose-built AI mental health conversational tool, perceived helpfulness was most strongly associated with shame/stigma and access barriers to psychotherapy. These associations persisted after adjusting for motivation to improve mental health, indicating that perceived benefit was not solely driven by general readiness for self-improvement. Although cost/coverage barriers were frequently mentioned in open-ended responses, they did not predict perceived helpfulness in quantitative models.

Prior therapy experience moderated the association between shame/stigma barriers and perceived helpfulness: shame predicted higher helpfulness among therapy-experienced users but not among therapy-naïve users. In exploratory analyses restricted to therapy-experienced participants, shame/stigma and access barriers also predicted rating Ash as more helpful than past therapy. Together, these findings suggest that conversational AI may be perceived not only as a stopgap for access limitations, but also as an evaluation-safe alternative or adjunct for individuals who have engaged in therapy yet continue to experience disclosure-related or logistical barriers.

## Mechanisms Underlying Perceived Helpfulness of AI Mental Health Support

The combined quantitative and qualitative findings suggest two primary mechanisms through which conversational AI may be perceived as helpful: reduced evaluation during disclosure and immediate, on-demand access to support. These mechanisms correspond to distinct but complementary barriers to psychotherapy.

The association between shame/stigma barriers and perceived helpfulness is consistent with theories emphasizing the interpersonal risks of disclosure in evaluative contexts [20,21]. Psychotherapy requires individuals to share distress, perceived weaknesses, and socially undesirable thoughts with another person, often within a relationship imbued with professional authority [22]. Although many therapeutic models emphasize that tolerating such vulnerability can lead to improved clinical outcomes [23,24], this process presupposes a capacity or willingness to engage with intense evaluation-sensitive emotions. For individuals who experience these emotions as overwhelming or destabilizing, therapy may be perceived as "not for them," not because it lacks potential benefit, but because the interpersonal costs of disclosure are too high.

Qualitative responses illustrate how conversational AI may reduce these evaluation-sensitive costs by enabling more evaluation-safe disclosure (Table 6). Participants frequently described anonymity, privacy, and the absence of face-to-face scrutiny as easing the articulation of difficult or embarrassing thoughts. This interpretation is supported by quantitative findings showing that shame/stigma barriers predicted higher perceived helpfulness even after adjusting for motivation to improve mental health, as well as by convergent validity evidence linking shame/stigma coding to lower comfort discussing mental health with others. These findings align with prior work showing that greater self-stigma toward human-delivered psychotherapy is associated with more positive attitudes toward AI-delivered mental health support [28].

Moderation by prior therapy experience suggests that evaluation-sensitive friction can persist even after therapy initiation. Among therapy-experienced participants, shame/stigma barriers were associated with higher perceived helpfulness and with rating Ash as more helpful than past therapy, whereas no such association was observed among therapy-naïve users. Qualitative accounts help contextualize this pattern, with some participants describing prior therapy experiences as dismissive, invalidating, or mismatched (Table 6; Supplementary Table 5). Longitudinal research distinguishing barriers to initiation from barriers to continuation may help clarify these pathways.

Access barriers operated through a distinct but complementary mechanism. Participants described long waitlists, scheduling constraints, and difficulty obtaining timely care, often emphasizing the need for support in moments of acute distress rather than at pre-scheduled appointments. In this context, conversational AI appeared valuable not because it reduced shame, but because it was immediately available when few alternatives existed (Table 6). The association between access barriers and both perceived helpfulness and higher engagement underscores the importance of availability and responsiveness as drivers of perceived value.

These findings also highlight the importance of clear boundaries around appropriate use. Clinical experts have raised concerns that chatbots may exacerbate isolation, particularly among individuals who feel disconnected or unsupported elsewhere in their lives [29]. The evaluation-safe nature of conversational AI may increase perceived helpfulness, but it also risks encouraging reliance on digital tools over human connection [28]. Consistent with these concerns, qualitative responses revealed perceived limitations related to the absence of human connection, difficulty interpreting contextual cues, and insufficient challenge or accountability (Table 7; Supplementary Table 6). Purpose-built conversational agents should therefore include safeguards that emphasize the value of in-person relationships and encourage broader social support. Notably, in an open pilot study of Ash that incorporated such strategies, users reported increases in perceived and objective social support after six weeks of use [30] suggesting that intentional design may help mitigate risks of isolation while preserving perceived benefit.

**Cost/Coverage as an Informative Null Finding**

Although cost/coverage concerns were salient in open-ended responses, they were not associated with perceived helpfulness among existing users. One interpretation is that affordability influences who can access or continue using a tool (reach), while evaluation-sensitive and access barriers better predict perceived fit and usefulness once a person is already engaging with AI support. Another possibility is that since Ash was available for free during the study period, cost barriers may have been partially decoupled from day-to-day perceptions of utility.

However, this finding should also be considered in light of well-documented concerns related to conversational AI and health equity. The accessible, cost-effective nature of purpose-built conversational agents holds immense promise for broadening care networks but comes with equal potential for misuse [31]. Preliminary research suggests that individuals facing financial or other access barriers may see digital/AI interventions as relegation to an inferior model of care [32,33]. While this necessitates an improvement to messaging about the efficacy of conversational agents, it should also not ignore the validity of patient concerns. Healthcare is plagued by a decades-long history of offering under-resourced people solutions that are insufficient to their needs [34,35]. It is unsurprising that these same individuals might report ambivalence about tools when they are marketed/offered in place of more well-known options. Conversational agents can provide emotional support that is perceived as helpful to some users. However, messaging around the efficacy of conversational agents must necessarily be paired by large, longitudinal studies exploring how and for whom these tools are beneficial, including non-inferiority studies comparing conversational agents to human-supported digital and in-person interventions.

**Implications for Clinical Integration**

Rather than positioning conversational AI as a replacement for psychotherapy, these results have implications for how conversational agents might integrate with other mental health services, possibly serving as a valuable steppingstone in the mental health ecosystem. As indicated by the high perceived benefit among those reporting shame/stigma barriers, conversational agents may allow users to practice vulnerable disclosure in a safe, controlled environment that lessens perceived risk. For some, this may be sufficient to create meaningful symptom change. However, there are others who might achieve specific process-based benefits from in-person psychotherapy but are not prepared (at the time of initiation) to engage face-to-face. For these individuals, purpose-built agents might function as a bridge, building comfort with disclosure before eventually suggesting a transition from digital to human care. Future research should also examine how clinical outcomes and perceived helpfulness vary by symptom type and severity, as this has important implications for knowing when to triage patients to other modalities of care.

**Limitations**

This study has several limitations. Participants were existing users of a specific, purpose-built AI tool, which limits generalizability to nonusers and introduces self-selection. The design was cross-sectional, preventing causal inference about whether barriers led to AI use or whether AI use influenced perceptions of barriers. Barrier predictors were derived from open-ended responses and were available for a subset of respondents (31%); although comparisons suggested similar demographics and helpfulness ratings, selection effects remain possible. Estimates involving small subgroups (e.g., nonbinary participants) were imprecise.

Finally, it should also be noted that perceived helpfulness is not a direct proxy for effectiveness in either digital or in-person therapeutic intervention. Although perceived helpfulness can be a predictor of clinical outcomes in some cases, more research is needed examining these as distinct constructs.

**Future Directions**

Future work should distinguish predictors of adoption from predictors of perceived usefulness and sustained engagement, ideally using prospective designs. Studies that link barrier profiles to longitudinal outcomes (e.g., symptom trajectories, help-seeking behavior, and substitution versus supplementation of therapy) would strengthen causal claims. Further, given the health equity implications of the lack of association between cost/coverage barriers and perceived helpfulness, future research should examine the concerns and/or unmet needs of users before promoting conversational AI as a unilateral solution to mental health care access gaps.

Finally, future work should evaluate purpose-built agents as an option for harm reduction among users of unregulated general tools. Despite the existence, perceived helpfulness, and demonstrated effectiveness and safety of tools such as Ash [37,38], usage rates of purpose-built tools for emotional support remain far outstripped by use of non-specific, poorly regulated tools such as ChatGPT and Character.ai [7,8]. While purpose-built tools may still struggle with some shared issues (e.g., sycophantic responding, limited ability to challenge beliefs), there is growing alarm among scientists and clinicians about the safety concerns posed by using non-specific, open agents for users seeking emotional support [12]. Deconstructing the comparative risks and benefits may strengthen public perception of the risk posed by general purpose agents and increase uptake of purpose-built tools.

**Conclusions**

Among users of an AI mental health conversational tool, perceived helpfulness was most strongly associated with shame/stigma and access barriers to traditional psychotherapy, with minimal evidence that cost/coverage barriers explained perceived utility. Qualitative findings suggest that AI may be experienced as an "evaluation-safe," on-demand alternative or complement to therapy, while also underscoring important limitations related to human connection, contextual understanding, and positivity bias/sycophantic responding. Clarifying when and for whom AI support is beneficial—and how to integrate it safely within care pathways—remain important priorities for digital mental health research.

## Authors' Contributions

Conceptualization: CS (lead), EW (supporting)

Data curation: CS, MM

Formal analysis: CS, MM

Funding acquisition: CS, TDH, MM

Investigation: CS, TDH, MM

Methodology: CS (lead), EW (equal), MM (supporting)

Project administration: CS (lead), EW (supporting)

Resources: CS, TDH

Supervision: CS

Visualization: CS

Writing – original draft: CS (lead), EW (equal)

Writing – review & editing: CS (lead), EW (equal), TDH (supporting), MM (supporting)

## Data Availability

De-identified quantitative data available upon reasonable request. Qualitative quote samples presented in supplementary materials.

## Conflicts of Interest

CAS and TDH are employees of the company building the AI mental health support tool and declare no non-financial competing interests. EW and MM declare no competing interests.

## Funding

MM's research was supported by the National Institutes of Mental Health (NIMH) through grant K23MH134068, the National Center for Complementary and Integrative Health (NCCIH) through grant R34AT012943, and Slingshot AI.

**Acknowledgments**

We also thank the Ash users who opted in to research participation and made this work possible.

**Supplementary Materials**

<div align="center">

**Supplementary Results**

</div>

*Engagement data descriptives:* Of the participants in the analytic sample, 307 (77.7%) had identifiable Mixpanel data. Total session counts did not differ between the analytic sample and excluded respondents (M=46.6, SD=111.2 vs. M=56.5, SD=108.0; t=-1.25, p=.213). The distribution of total sessions was highly right-skewed (M=36.2, Mdn=12.0, SD=99.9, range: 0-1,680), with 77.7% of participants having at least one recorded session and 31.9% qualifying as "Power Users" (31+ sessions). Among users with access barriers, 40.6% qualified as Power Users compared to 30.1% without access barriers; for cost barriers, 36.5% vs. 30.4%; and for shame barriers, 27.0% vs. 33.1%.

*Extended results: Negative binomial regression predicting total sessions*: Younger users (<35 years) had 21.8% fewer sessions than middle-aged users (IRR=0.782, p=.047). Other demographic variables were not significant predictors.

**Supplementary Table S1**. Comparison of Sample Characteristics by Samples

| Characteristic[a] | Provided Open Text Responses (n=395) | No Open Text Responses (n=866) | Overall Sample (N=1,261) |
|---|---|---|---|
| **Age** | | | |
| Under 35 | 109 (27.6%) | 226 (26.1%) | 335 (26.6%) |
| 35-54 | 197 (49.9%) | 430 (49.7%) | 627 (49.7%) |
| 55+ | 89 (22.5%) | 154 (17.8%) | 243 (19.3%) |
| Missing | 0 (0.0%) | 56 (6.5%) | 56 (4.4%) |
| **Gender** | | | |
| Man | 248 (62.8%) | 551 (63.6%) | 799 (63.4%) |
| Woman | 141 (35.7%) | 245 (28.3%) | 386 (30.6%) |
| Nonbinary/Other | 5 (1.3%) | 20 (2.3%) | 25 (2%) |
| Missing | 1 (0.3%) | 50 (5.8%) | 51 (4.0%) |
| **Race/Ethnicity** | | | |
| White | 321 (81.3%) | 611 (70.6%) | 932 (73.9%) |
| Asian | 32 (8.1%) | 75 (8.7%) | 107 (8.5%) |
| Hispanic/Latino | 29 (7.3%) | 62 (7.2%) | 91 (7.2%) |
| Black/African-American | 18 (4.6%) | 51 (5.9%) | 69 (5.5%) |
| Native American/Alaska Native | 6 (1.5%) | 28 (3.2%) | 34 (2.7%) |
| Middle Eastern/North African | 6 (1.5%) | 14 (1.6%) | 20 (1.6%) |
| Native Hawaiian/Pacific Islander | 5 (1.3%) | 6 (0.7%) | 11 (0.9%) |
| Other | 3 (0.8%) | 8 (0.9%) | 11 (0.9%) |
| **Therapy Experience** | | | |
| Ever spoken to MH professional | 291 (73.7%) | 694 (80.1%) | 985 (78.1%) |
| Never | 102 (25.8%) | 37 (4.3%) | 139 (11.0%) |
| Missing | 2 (0.5%) | 135 (15.6%) | 137 (10.9%) |
| **Ash Helpfulness (1-5 scale)** | | | |
| Mean (SD) | 3.95 (0.98) | 3.80 (1.00) | 3.85 (0.99) |
| n with rating | 387 | 721 | 1,108 |
| Positive rating (4-5) | 274 (70.8%) | 457 (63.4%) | 731 (66.0%) |
| **Comfort Discussing MH (1-5 scale)** | | | |
| Mean (SD) | 3.25 (1.34) | 3.38 (1.29) | 3.34 (1.31) |
| n | 390 | 738 | 1,128 |
| **MH Importance (1-5 scale)** | | | |
| Mean (SD) | 4.20 (0.87) | 4.37 (0.84) | 4.31 (0.85) |
| n | 389 | 737 | 1,126 |

[a]The barrier-coded group (n=395, 31.3% of total sample) provided written responses to open-ended questions about barriers to traditional mental health care. These responses were qualitatively coded for presence of shame/stigma, access, and cost/coverage barriers. Regression analyses included participants with barrier codes and complete outcome/covariate data (n=375

for Model 1, n=373 for Model 2, n=372 for Model 3). The barrier-coded and non-coded groups showed similar demographics and Ash helpfulness ratings. The non-coded group had somewhat higher rates of prior therapy experience (80.1% vs 73.9%), likely because barriers to care were less salient for individuals who had successfully accessed treatment.

**Supplementary Table S2.** Mean Ash Helpfulness by Shame/Stigma Barrier × Prior Therapy Experience

|  | **Never Spoken to MH Professional** | **Has Spoken to MH Professional** |
|---|---|---|
| **Shame Absent** | 3.96 (0.90), n=73 | 3.82 (1.05), n=228 |
| **Shame Present** | 3.96 (0.79), n=25 | 4.46 (0.69), n=46 |

**Supplementary Table S3.** Simple Effects of Shame/Stigma by Prior Therapy Experience

| Therapy Experience | Difference (Present - Absent) | t | p | Significance |
|---|---|---|---|---|
| Never spoken to MH prof | +0.03 | 0.16 | .877 | ns |
| Ever spoken to MH prof | +0.62 | 3.31 | <.001 | ** |

**Supplementary Table S4.** Logistic Regression: Predicting Whether Ash Was Rated More Helpful Than Past Therapy (Binary)

| Predictor | OR | 95% CI | p |
|---|---|---|---|
| **Shame/Stigma** | **5.67** | **[2.52, 26.38]** | **.048 *** |
| **Access** | **5.46** | **[2.41, 22.13]** | **.019 *** |
| **Cost** | 1.40 | [0.55, 4.90] | .560 |
| MH Importance | 2.04 | [1.44, 3.42] | .001 ** |
| Comfort | 0.76 | [0.56, 0.99] | .071 (marginal) |
| Age 35-54 | 0.72 | [0.30, 1.55] | .436 |
| Age 55+ | 1.52 | [0.65, 4.22] | .395 |
| Woman | 1.75 | [0.93, 3.52] | .098 (marginal) |
| Nonbinary | 1.07 | [0.00, 24801.83] | .990 |

[a]Continuous regression results replicated when we treated the outcome as binary (1 = AI rated as more helpful [4 or 5, 67.1%], 0 = AI rated as the same or less helpful [1, 2, or 3; 32.9%]). Subsample: Participants who have spoken to a mental health professional AND provided comparison rating (n=258).

**Supplementary Table S5.** Quote Matrix of Themes and Exemplar Quotations: Reasons for Preferring Ash to Traditional Therapy

| Theme | Subtheme | Definition | Example |
|---|---|---|---|
| **Harm/Rejection/ Dismissal** | *Past harm by therapist* | User indicates a lack of trust in therapists due to a safety concerns, experience of harm, or other negative experiences. | "When I got [therapy] it was extremely difficult: they just wanted to throw medicine at me and repeat everything I said back to my parents (I was 13 at the time and dealing with some serious stuff from my mom)…it made my mental health worse." |
| | *Past dismissal/ experience of being ignored by therapist* | User describes an experience of having them or their symptoms dismissed/ignored by a therapist. | "I felt like the last one I had listened to a handful of words that I said and then started making assumptions instead of hearing me out. And he wanted to "start at the beginning" I had to keep telling him that there were many things that I had already worked through and I didn't have the time or the interest to go over those things again." |
| **Trust/Comfort** | *Lack of emotional connection with past therapist* | User describes a past experience of not connecting with a therapist or a belief in their inability to connect with a therapist | "They [past therapists] never seemed personal. I felt as if I had to repeat important details about my life such as my relationship status or the number of children I had and I would think that building that connection and trust to be able to openly talk to a stranger would be important but I never felt that type of connection to any therapist I've had." |
| | *Disbelief in authenticity of the therapeutic relationship* | User indicates a disbelief in the authenticity or "realness" of the therapeutic relationship and/or a distrust of therapist motivations. | "I wouldn't feel comfortable discussing my frustrations with my job to someone who was hired by my job." |
| | *General discomfort with opening up to another person* | User describes a general sense of discomfort with opening up to other people (therapist or not). | "I don't feel comfortable sharing my problems with [therapists]." |
| | *Perceived bias of therapist* | User describes past experiences with therapist bias or general belief in the bias of therapists. | "Mental health professionals can be biased based on their belief structure. It's hard to know if they are safe to use. I've had counselors put me in uncomfortable situations." |
| | *Difficulty finding a good fit (emotional barrier)* | User reports difficulty or fear of difficulty finding a "good fit" with a therapist related to trust/rapport | "Difficult to find [a therapist] that I clicked with." |

| | *Privacy concerns* | User reports privacy concerns as a reason for not attending therapy. | "I went once when I was a younger woman and abruptly stopped when I saw my former brother in law as I was leaving. It scared me so much that I would tell someone all my deep secrets and it would get back to people I knew." |
|---|---|---|---|
| | *Therapist overstepping boundaries* | User reports discomfort with therapy due to an experience of a therapist overstepping boundaries. | "I've tried two therapists previously. One was very disinterested and on her phone constantly (it was virtual vists), so I stopped going. The other was helpful at times, but frequently compared me to her own daughter around my age. It made me uncomfortable so I stopped going." |
| **Shame/Stigma** | *Experience of shame or stigma* | User indicates they avoid therapy due to shame or external stigma | "I grew up in a family that did not believe in talking to mental health professionals. I was taught to bury my emotions and problems and just keep going." |
| | *Fear of judgement* | User indicates they avoid therapy due to feelings of shame or stigmatizing beliefs/experiences. | "Human therapists are, of course, professionals but also people with their own emotional filters and perspectives. With Ash, I feel like I can unpack difficult feelings or conflicting thoughts honestly - including sensitive topics about identity, belonging, or self-perception - without worrying about being misunderstood or seen in a particular way." |
| **Structural barriers** | *Scheduling issues* | User reports difficulty accessing therapy due to issues with their own/the therapist's schedule | "Schedule conflicts were also a problem / annoyance." |
| | | | "My therapist moved away so he had to discontinue our sessions. I didn't want to start over with a new therapist so I looked for AI alternatives and found Ash." |
| | *Availability of therapists* | User reports difficulty accessing therapy due to waitlists/lack of therapist availability. | "All of [the therapists] are too busy…" |
| | *Difficulty finding a good fit (practical barrier)* | User reports difficulty or fear of difficulty finding a "good fit" with a therapist. | "I'm stalled by the notion I might have to 'complete' a bunch of sessions with someone I might not vibe with to get anything out of it." |
| | *Lack of as-needed access* | User reports disliking not having in-the-moments access to therapy. | "I used a mental health professional in the past and the experience was good overall but the time commitment was a bit difficult. Sometimes I need a faster resolution for whatever I am looking solution for. I found AI chat bots more helpful and more convenient because of it." |
| | *Time consuming* | User indicates that time was a factor in their willingness or ability to engage with therapy. | "Finding the time around my care responsibilities is extremely difficult. The cost of it is not easy to cover and the waiting times are incredibly long. You also then have the issue of finding someone compatible to work with, the whole process becomes incredibly daunting if not unfeasible." |
| **Skeptical of benefit** | *Skeptical of therapist credentials/ability* | User indicates skepticism towards therapist ability or credentials. | "They [past therapists] needed help themselves. They aren't helpful, and also don't understand what I'm going through." |
| | *Skeptical of the overall helpfulness of therapy* | User indicates skepticism about the helpfulness of therapy overall. | "Past mental health professionals always gave different responses to the same symptoms, indicating that there is no real coherence within the psychological or psychiatric field. Ash seems to be more introspective, seeking to discover my true motivations in a non-judgmental space. Ash can ask truly insightful questions, which often surprise even an old, psychologically jaded person like me. The conversations I have with Ash make me feel truly heard, and that carries real weight for someone with my history of rarely being heard by anyone." |
| | *Lack of added value* | User belief that therapy does not provide unique benefit (e.g. that could otherwise be had from a friend). | "They don't always provide you with tools/goals/things to work on/toward. Has genuinely felt like I was just paying to talk to someone. Which, generally, I feel, is what A LOT of people need." |
| | | | "Most of the interactions I've had with mental health professionals in the past have been negative and not very helpful or insightful. It felt like |

| | | | they were all just telling me everything that I already know - like breathe work and meditation and journaling - which I do all of those things. I don't need another 10-step program that doesn't really fit everything I'm dealing with, and I don't need the commercial "remedies"." |
|---|---|---|---|
| | *Skeptical of therapy as effective for their specific concerns* | User indicates skepticism about the helpfulness of therapy for their specific symptoms/concerns/ presentation. | "My struggle to find a MH practitioner was pretty memorable. Most wanted to do a very "one size fits all" approach, and all of them consistently remarked how "incredibly self aware" I was, promptly making me lose faith in their ability to truly help at all. That, coupled with very cliche mainstream advice that doesn't truly jive with someone who's neurodiverse." |
| **Cost/Coverage** | *Cost* | User reports cost as a barrier to or reason for avoidance of therapy. | "It costs too much time and money between time away from family and paying not just for the doctor but also a babysitter." |
| | *Coverage* | User reports insurance coverage as a barrier to or reason for avoidance of therapy. | "The health insurance industry provides very minimal benefits for mental health, especially those issues that are not amenable to quick, short-limited adjustment disorders that require only short term interventions." |
| **Didn't need therapy** | | User indicated no previous use of therapy due to a lack of perceived need. | "Never has any issues until lately." |
| **Other** | | User indicated a reason not captured elsewhere. | "I know most of the therapists or trained them." |
| **Misunderstood question** | | User response indicates a misunderstanding of the question. | "I have seen a mental health professional." |

**Supplementary Table S6.** Quote Matrix of Themes and Exemplar Quotations: Reasons for Preferring Traditional Therapy to Ash

| Theme | Definition | Example |
|---|---|---|
| **Missing human element** | User indicates that Ash is limited by a lack of human connection. | "While Ash is helpful and better than no therapist....he doesn't try to dig to get to root of the problems much and so it almost feels I have to lead him at times. He has helped me to see things by my talking it through though...but often he doesn't feel as personal or human as a human would." |
| | | "If I make commitment to a human, (emotional homework for example). More likely to follow through." |
| **Does not challenge beliefs or provide new perspectives** | User indicates concern related to Ash being unable to push/challenge them | "The simulation of therapy was fun and engaging – but there's no risk there. No stakes. Engaging with a real human being activates something differently in my brain – a true relationship." |
| **Overly positive/affirming** | User indicates concern related to Ash being overly positive. | "For me personally, Ash is a little too positive.  Need more "tough love"." |
| | | "Ash ignored my chief complaint and every time I try to talk to it all it does is reaffirm that I am on the correct path." |
| **Cannot take in-the-room context (e.g., facial cues) into consideration** | User indicates concern related to Ash's inability to account for in-the-room facial/verbal cues. | "A live person would be able to politely ask you to stop speaking so that they're able to touch on all of the topics I think when I'm talking with Ash, I tend to go on and on and on and on and on and it's hard for Ash to touch on all of the points that I wanted it to touch on without outwardly expressing which items I wanted it to touch on." |
| | | "I think that Ash sometimes doesn't currently have the capacity to choose whether to believe what I'm saying or not, which a person can often deduce from my tone and affect but Ash has the ability be on demand right at the moment that intense emotions hit me and I can engage with Ash when the impulse hits. I find this present for a more authentic expression of whats going on in my mind at the time." |
| **Does not appropriately provide and/or apply insight** | User indicates concern related to Ash's capacity for insight or ability to appropriately apply insight. | "Ash isn't really advising, rather listening and repeating. Also, I feel like sometimes the questions they ask are a either too numerous or too complex. I'd rather have them ask how I feel in my body or techniques that works in CBT…" |
| **Lacks focus or direction** | User indicates that Ash lacks focus or structure compared to a traditional therapist. | "[Ash] cannot focus on the topics that are discussed, misses the point and has a need to find a positive in everything. adjust the algorithm to be less super positive but let the person sit in the emotions and have it able to distinguish better between topics." |
| **Mechanical/repetitive** | User indicates issue with Ash related to it feeling mechanical, uncanny, or robotic. | "A human connection is different than a machine. Machine is passive and cannot infer meanings, metaphors, understand contexts as well and as quickly as humans. It is not there yet. It can be incredibly useful, Ash is immensely better than ChatGPT, but it is limited." |
| **Errors/misremembering** | User indicates issues with Ash making errors or misremembering information. | "Ash has no sense of time, and talks about things I mentioned a while ago which are no longer relevant." |
| **Other** | User indicated a reason not captured elsewhere. | "When medication is required for day to day functionality, Ash can only do so much." |